# Percentile-Focused Regression Method for Applied Data with Irregular Error Structures

Elsayed A. H. Elamir
Department of Management and Marketing, College of Business Administration, Kingdom of Bahrain
Email: shabib@uob.edu.bh

## Abstract

Irregular errors such as heteroscedasticity and non-normality remain major challenges in linear modelling. These issues often lead to biased inference and unreliable measures of uncertainty. Classical remedies, such as robust standard errors and weighted least squares, only partially address the problem and may fail when heteroscedasticity interacts with skewness or nonlinear mean structures. To address this, we propose a two-stage cumulative distribution function-based (CDF-based) beta regression framework that models the full conditional distribution of the response. The approach first transforms the outcome using a smoothed empirical CDF and then fits a flexible beta regression, allowing heteroscedasticity and non-normality to be handled naturally through the mean–precision structure of the beta distribution. Predictions are mapped back to the original scale via the empirical quantile function that preserving interpretability. A comprehensive Monte Carlo study shows that the proposed method consistently achieves good distributional accuracy and well-calibrated prediction intervals compared with OLS, WLS, GLS. Application to the concrete compressive strength dataset demonstrates its stability and practical advantages.

ORCID: 0000-0002-9430-072X

# 1 Introduction

Ordinary Least Squares (OLS) linear regression remains a foundational tool in empirical research across many fields such as economics, social sciences and epidemiology. See; Wooldridge (2020). While OLS coefficient estimates remain unbiased even when certain classical assumptions are violated, the validity of statistical inference such as confidence intervals, and p-values, depends critically on two key assumptions: homoscedasticity and normality of residuals (Fox, 2021). Recent studies have refined the understanding of when and how these violations matter, especially in modern contexts involving high-dimensional data, complex sampling designs, and non-standard error structures (Amado et al., 2025; Pinzon, 2022). Studying heteroscedasticity and non-normality is fundamental for ensuring the validity, efficiency, and credibility of empirical research. While OLS remains robust to assumption violations for point estimation, valid inference requires careful attention to heteroscedasticity and non-normality where the standard OLS variance-covariance estimator becomes inconsistent, leading to biased standard errors. Consequently, $t$-statistics, $F$-tests, and confidence intervals may be severely misleading (Pinzon, 2022 and Amado et al., 2025).

Simulation evidence shows that problems with statistical inference are influenced not only by the overall sample size, but also by the ratio of observations to the number of regressors ($n/k$). When this ratio is small (for example, less than 50), even heteroscedasticity-consistent (HC) standard errors can perform poorly and fail to correct the bias adequately. The issue becomes more serious in the presence of high-leverage observations, which may further distort HC estimators, particularly HC1, making standard errors unreliable (Amado et al., 2025, Heller, 2024 and White, 1980).

Although the central limit theorem offers protection in very large samples, many applied studies work with moderate or small datasets, such as clinical trials or macroeconomic analyses. In these settings, non-normal errors can distort the sampling distribution of estimators, leading to misleading p-values and unreliable inference, as highlighted in the literature (e.g., Gelman et al., 2020; MacKinnon, 2023; White, 1980).

Recent work emphasizes that extreme non-normality, particularly heavy tails or severe skewness, can distort inference even in moderate samples, especially when combined with heteroscedasticity or high leverage. Heteroscedasticity and non-normality often co-occur and can amplify inference problems. For instance, heavy-tailed errors combined with heteroscedasticity can cause HC standard errors to under-cover confidence intervals. Also, outliers may induce both non-constant variance and skewness; robust regression methods that downweigh leverage points address both issues simultaneously. Under heteroscedasticity, OLS is no longer BLUE although it remains unbiased, but it has a larger variance than necessary, see, (Amado et al., 2025 and Arai et al. 2024).

Transformations are mathematical modifications applied to variables (usually the dependent variable, but sometimes predictors) to help a linear regression model better satisfy its classical assumptions. Examples include Log, square-root, or inverse transformations are useful for right-skewed outcomes. Box-Cox transformations can normalize residuals, but optimal $\lambda$ should be selected via cross-validation to avoid overfitting (Farrar et al., 2025). These traditional methods or generalized least squares (GLS) address only parts of the problem and often introduce new weaknesses. Transformations alter the scale of the response, complicating interpretation and introducing bias through back-transformation. Moment-based corrections assume the researcher can correctly guess the variance structure, an assumption that breaks down quickly when heteroscedasticity interacts with non-normality or nonlinear mean patterns (Fox and Weisberg, 2019, Godfrey and Orme, 1999 and Goffrey et al., 2006).
In this regard Beta regression has emerged as the standard methodology for modeling continuous dependent variables restricted to the standard unit interval $(0, 1)$, such as rates, and proportions. Unlike ordinary least squares, beta regression explicitly accounts for the heteroscedasticity and non-normality inherent in the data. As noted by Ferrari and Cribari-Neto (2004), the variance of a beta-distributed variable is a function of its mean.
Therefore, in this study we propose a new approach called cumulative distribution function beta regression (CDF-based beta regression) that deals with the limitations of classical regression in case of heteroscedasticity and non-normality. Instead of trying to fix violations of assumptions, this method remodels the problem by transforming the response into its empirical distribution value and modelling it using a flexible beta family. This accomplishes three things that traditional methods cannot achieve simultaneously: the method models the entire conditional distribution rather than the mean or variance, heteroscedasticity is also absorbed naturally in beta regression that eliminating the need to guess variance functions or specify weighting schemes and effects are interpreted through percentile shifts by returning to the real data scale using the empirical quantile function. These preserves meaning for applied researchers, avoiding the interpretational distortions caused by log, Box–Cox, or reciprocal transformations (Benoit, 2011 and Atkinson et al., 2021).
While the classical beta regression requires the raw response to lie in (0,1), the proposed approach can model any continuous distribution by creating a beta-compatible response via the empirical CDF transformation even when heavy-tailed, skewed, or bounded on an arbitrary scale. In addition, the proposed approach models the full conditional distribution in a single likelihood that guarantee coherent predictions across percentiles while the quantile regression estimates conditional quantiles one at a time and does not enforce consistency across quantiles,
This study is organized as follows. Section 2 reviews some classical linear modelling such as OLS, WLS and GLS. Section 3 introduces the theoretical foundations of beta regression, the empirical CDF and quantile transformation. Section 4 presents the proposed two-stage CDF-based beta regression framework and develops the bootstrap procedure for inference and prediction interval construction. Section 5 describes the design and reports the results of the Monte-Carlo simulation study. Section 6 applies the CDF-based beta regression method to the concrete compressive strength dataset. Finally, Section 7 offers concluding remarks.

## 2 Least squares

Ordinary least squares (OLS) is the foundational estimation method for linear regression models of the form

$$y = \mathrm{X}\beta + \epsilon$$

$y$ $(n \times 1)$ is vector of dependent variable, $X$ $(n \times p)$ is a matrix of predictors, $\beta$ $(p \times 1)$ is a vector of unknown parameters and $\epsilon$ $(p \times 1)$ a vector of errors. OLS estimates the parameter vector

$$\hat{\beta} = \arg\min_{\beta} \sum_{i=1}^{n} (y_i - x_i^{\top}\beta)^2.$$

The solution is

$$\hat{\beta} = (X^{\mathrm{T}}X)^{-1}X^{\mathrm{T}}y$$

provided $X^{\mathrm{T}}X$is invertible. Under the classical assumptions, OLS is unbiased, consistent, efficient (minimum variance among all linear unbiased estimators — the BLUE property). These assumptions are linearity in parameters, no perfect multicollinearity, zero conditional mean ($\mathrm{E}(\, \varepsilon_i \mid X \,) = 0$), homoscedasticity ($\mathrm{Var}(\, \varepsilon_i \mid X \,) = \sigma^2$), normality of errors ($\varepsilon_i \sim N(0, \sigma^2)$) and no autocorrelation ($\mathrm{Cov}(\varepsilon_i, \varepsilon_j) = 0, i \neq j$) (Wooldridge, 2020). Moreover, under these assumptions, the covariance matrix is

$$Cov(\hat{\beta}) = \sigma^2 \left(\acute{X}X\right)^{-1}$$

Furthermore, $\hat{\sigma}^2 = (n-p)^{-1}\acute{e}e$ is an unbiased estimator of $\sigma^2$ and $e$ is error vector.

When residual variance differs across observations, i.e., heteroscedasticity, OLS remains unbiased but becomes inefficient, and its standard errors become unreliable (Arai et al. 2024, Gujarati, 2018). The weighted least squares address this by assigning weights inversely proportional to the error variance

$$Y = X\beta + \varepsilon, \varepsilon \sim N(0, \Sigma),$$

where $\Sigma = \mathrm{diag}(\sigma_1^2, \ldots, \sigma_n^2)$ has unequal variances. Each weight is $w_i = 1/\sigma_i^2$. Then WLS estimator is

$$\hat{\beta}_{WLS} = (X^{\top}WX)^{-1}X^{\top}WY.$$

This minimizes the weighted sum of squared residuals

$$\hat{\beta}_{WLS} = \arg\min_{\beta} \sum_{i=1}^{n} w_i \, (y_i - x_i^{\top}\beta)^2.$$

WLS is essentially OLS applied after rescaling observations according to their information content. Observations with smaller variance receive larger weight, because they contain more reliable information. WLS is a special case of generalized least squares (GLS) when the error covariance matrix is diagonal. GLS generalizes the approach to allow for heteroscedasticity and autocorrelation where Σ may have non-zero off-diagonal elements (correlated errors). Weighted least squares becomes GLS when the entire covariance matrix Σ is used instead of only its diagonal GLS solves

$$\hat{\beta}_{GLS} = (X^{\top}\Sigma^{-1}X)^{-1}X^{\top}\Sigma^{-1}Y$$

produces a transformed model with homoscedastic, uncorrelated errors, to which OLS is then applied (Green, 2012, Ding, 2024 and Gujarati, 2018).

# 3 Beta regression and distribution function

Beta regression is a statistical modelling technique designed for response variables bounded in the interval (0, 1)—such as proportions, rates and percentages. Unlike linear regression, which assumes normality and homoscedasticity, beta regression directly models data that are typically heteroscedastic and skewed. Unlike OLS, which assumes normality and constant variance, beta regression is built around the beta distribution, which is naturally flexible in shape (U-shaped, J-shaped, symmetric, or skewed), Heteroscedastic (variance depends on the mean) and appropriate for bounded responses (Daimon, 2008 and Ali et al., 2023).

Ferrari and Cribari-Neto (2004) proposed a parameterization using mean $E(Y) = \mu, 0 < \mu < 1$ and precision parameter $\phi > 0$. Under this parameterization, if

$$Y \sim \text{Beta}(\mu, \phi),$$

its density is

$$f(y; \mu, \phi) = \frac{\Gamma(\phi)}{\Gamma(\mu\phi)\Gamma\big((1-\mu)\phi\big)}\, y^{\mu\phi-1}(1-y)^{(1-\mu)\phi-1}, \quad 0 < y < 1.$$

The variance is $\text{Var}(Y) = \mu_i(1-\mu_i)/(1+\phi_i)$, showing directly that variance depends on the mean that inherent heteroscedasticity. A standard beta regression has two models (the mean and dispersion models). The mean model is given by

$$g(\mu_i) = \text{X}\beta$$

$\mu_i$is the mean of the beta-distributed response, $g(\cdot)$is a link function (typically logit, but probit, cloglog can also be used), $X$ is a covariate and $\beta$ is the regression coefficients. The precision model is

$$h(\phi_i) = Z\gamma$$

$\phi$ controls precision (inverse dispersion), $\phi$ may be constant or modelled with covariates, $Z$ is covariates and $\gamma$ is the coefficients.

Given a set of observations $y_i \in (0,1)$, the log-likelihood function of the beta regression model under the mean–precision parameterization $(\mu_i, \phi_i)$is written as

$$\ell(\beta, \gamma) = \sum_{i=1}^{n} [\log\, \Gamma(\phi_i) - \log\, \Gamma(\mu_i\phi_i) - \log\, \Gamma((1-\mu_i)\phi_i) + (\mu_i\phi_i - 1)\log\, y_i + ((1 - \mu_i)\phi_i - 1)\log\,(1-y_i)]\,,$$

a formulation that follows the standard parameterization proposed by (Ferrari and Cribari-Neto, 2004). In this framework, the conditional mean $\mu_i$is modeled using a link function such that $\mu_i = g^{-1}(X_i^{\top}\beta)$. With the logit link being the most common choice due to its interpretability and strict mapping into the unit interval (Cribari-Neto and Zeileis, 2010). The precision parameter $\phi_i$, which governs the dispersion structure of the response variable, is specified through a separate submodel of the form $\phi_i = h^{-1}\big(Z_i^{\top}\gamma\big)$. Typically using a logarithmic link to ensure positivity. The dual-model formulation enables beta regression to directly account for heteroscedasticity, as the conditional variance of a beta-distributed outcome is $\text{Var}(Y_i) = \mu_i(1-\mu_i)/(1+\phi_i)$, inherently dependent on both the mean and the precision (Zeileis et al., 2020).

Maximum likelihood estimation of $\beta$and $\gamma$ requires numerical optimization since closed-form solutions do not exist for this likelihood function. Optimization routines such as Newton–Raphson, Fisher scoring, and quasi-Newton algorithms are typically employed in modern implementations (Amado et al., 2025). These iterative techniques update parameter values until convergence and provide consistent and asymptotically efficient estimates under standard regularity conditions, making them widely used in contemporary applied beta regression models (Zeileis et al. 2020) and (Maluf, 2025). The package “betareg” is used for all computations in this study ( R Core Team, 2026).

The cumulative distribution function (CDF), denoted as $F_Y(y)$, is the primary tool used to describe the behavior of a random variable $Y$ as

$$F_Y(y) = P(Y \leq y)$$

It represents the probability that the random variable Y takes a value less than or equal to some specific number y. This relationship is definitive where the distribution function completely characterizes the random Variable. The Quantile Function (QF) denoted as ($Q_Y$ or $F_Y^{-1}$) is the inverse of the distribution function

$$Q_Y(p) = \inf\{y: F_Y(y) \geq p\}$$

$p \in (0,1)$ is the probability and if $F_Y$is continuous and strictly increasing, they are perfect inverses

$$F_Y\big(Q_Y(p)\big) = p \text{ and } Q_Y\big(F_Y(y)\big) = y$$

The empirical distribution function (EDF) is used to approximate the cumulative distribution function (CDF) of an unknown distribution based solely on observed data (Serfling, 1980). From (Elamir, 2025), the FC-Hermite approach $\left(F_{C,i}(y)\right)$ is used where it is simple and a smoothed empirical distribution function estimator, based on Hermite-style interpolation principles, that produces stable, strictly-bounded, nearly continuous distribution function values and better suited for modelling purposes in beta regression as it preserves ordering while avoiding boundary issues and stepwise jumps

$$F_{C,i}(y) = \begin{cases} \dfrac{1}{n}, & i = 1 \\ \dfrac{2i-1}{2n}, & i = 2, \dots, n-1 \\ \dfrac{n-1}{n}, & i = n \end{cases}$$

With respect to quantile function, the Harrell–Davis estimator, which defines the quantile as a weighted average of all order statistics

$$\hat{Q}_{HD}(p) = \sum_{i=1}^{n} w_i\,(p) Y_{(i)},$$

where weights $w_i(p), p \in (0,1)$ arise from a Beta distribution

$$w_i(p) = B\left(\frac{i}{n};\ (n+1)p, (n+1)(1-p)\right) - B\left(\frac{i-1}{n};\ (n+1)p, (n+1)(1-p)\right),$$

Note that the Harrell-Davis (H-D) estimator, implemented as “hdquantile ()” function in “Hmisc” package in R-software (R Core Team, 2026) and (Harrell and Davis, 1982).
The H-D quantile estimator uses weighted averages of all order statistics, so its variance typically satisfies

$$\text{Var}\,(\hat{Q}_{HD}(p)) = w^{\mathrm{T}} \Sigma w,$$

$w$ weights depending on $p$, $\Sigma$ covariance matrix of the order statistics. Note that if the sample quantile is used the asymptotic variance is

$$\text{Var}\,(\hat{Q}_n(p)) \approx \frac{p(1-p)}{n\,f(Q(p))^2}$$

See, Serfling (1980).

## 4 Two-Stage CDF-based beta regression

Instead of relying on classical transformations such as the Box–Cox, reciprocal, or logarithmic transformations, the proposed approach models the entire distribution of the response variable through a cumulative distribution function (CDF) transformation**.** In the first stage, each observed response $Y_i$is transformed into its empirical distribution value

$$U_i = F_{C,i}(Y_i),$$

where $F_{C,i}(\cdot)$denotes a smoothed empirical distribution estimator. This transformation maps the response onto the unit interval while preserving its rank structure and eliminating scale-specific distortions associated with classical transformations (e.g., log or Box–Cox). Importantly, the empirical CDF is estimated from data, and therefore $U_i$ contains the usual sampling variability inherent in nonparametric distribution estimators.

While this approach introduces additional perturbation, the smoothed empirical CDF is consistent at rate $n^{-1/2}$(Theorem A.1 in Appendix). The perturbation

$$U_i = F_{C,i}(Y_i) + O_p(n^{-1/2}),$$

enters the likelihood only through a second-order contribution. Consequently, the empirical CDF transformation does not alter the first-order properties of the subsequent likelihood-based estimator. In the second stage, the transformed responses are assumed to follows a beta distribution parameterized by a mean $\mu_i \in (0,1)$ and a precision parameter $\phi > 0$, such that

$$U_i \sim \text{Beta}(\mu_i, \phi).$$

Under this parameterization, $E[U_i] = \mu_i$ denotes the expected CDF position, and the variance of the transformed response is given by $\text{Var}(U_i) = \mu_i(1-\mu_i)/(1+\phi)$, demonstrating that the conditional variance is inherently mean-dependent, a key feature of heteroscedastic data. Building on this representation, the framework specifies both a mean model, linking $\mu_i$to covariates through an appropriate link function, and a dispersion (precision) model, governing the behaviour of $\phi$ across observations. The mean model is

$$g(\mu_i) = \eta_i = \text{X}\beta$$

and the dispersion is

$$h(\phi_i) = Z_i^\top \gamma,$$

$g(.)$ link function, $\eta_i$ linear predictor, $\beta$ regression coefficients, $X$ covariates, $Z_i$ covariate or different set and $\gamma$ precision.

Model parameters $(\beta, \gamma)$ are obtained via maximum likelihood. Because the perturbation in CDF is asymptotically negligible relative to the $O(n)$ information in the beta likelihood, the estimator shares the same first-order asymptotic normality as the classical beta regression MLE (Theorem A.2 in Appendix). Therefore, the two-stage estimation procedure remains well-behaved, and no instability arises from the CDF transformation. The transformation effectively redefines the response scale without compromising likelihood-based estimation.

## 4.1 Mean model

The proposed mean model for two stage CDF-based beta regression is

$$g(\mu_i) = \eta_i = \mathrm{X}\beta$$

A variety of link functions such as the logit and probit can be employed depending on the distributional characteristics of the response variable and the modelling objectives. Among these, the logit link is the most widely used due to its interpretability. Therefore, in this study, we adopt the logit link function as the canonical choice for modelling the mean structure as

$$\mathrm{logit}(\mu_i) = \log\left(\frac{\mu_i}{1-\mu_i}\right) = \eta_i = \mathrm{X}\beta$$

To estimate or predict the dependent variable $y$, the quantile function is used as

$$\hat{y}_i = \hat{Q}_Y(\hat{\mu}_i)$$

$\hat{\mu}_i$ is the estimated or predicted cumulative distribution function that obtained from beta regression. A desirable feature of any regression methodology is that model components remain interpretable to applied researchers. The CDF-based beta regression approach offers two complementary modes of interpretation. The interpretation on the distribution (percentile) scale, and interpretation on the original response scale via the quantile function. This dual view provides a level of interpretability that is difficult to achieve with classical transformations or variance-correction methods. The change in a coefficient $\beta_j$describes how covariate $X_j$shifts the expected percentile position of the response variable. A positive $\beta_j$increases $\mu_i$, meaning the observation moves toward higher percentiles of the distribution; a negative coefficient shifts it toward lower percentiles. Since the conditional mean is given by $\mu_i = e^{\eta_i}/(1+e^{\eta_i})$, its marginal effect with respect to covariate $X_j$is

$$\frac{\partial\mu_i}{\partial X_j} = \beta_j\,\mu_i(1-\mu_i).$$

This expression shows that the influence of $X_j$on the mean is inherently nonlinear and depends on the current value of $\mu_i$. Because $\mu_i(1-\mu_i)$ varies across observations, the marginal effect is not constant, reflecting the heterogeneity in variance characteristic of beta regression models. While the model is fitted on the CDF scale, predictions are mapped back to the original outcome using

$$\hat{y}_i = Q_Y(\hat{\mu}_i),$$

This step yields direct interpretability on the original units. The fitted value $\hat{y}_i$is the predicted $\hat{\mu}_i$-th quantile of the response. For example, if $\hat{\mu}_i = 0.75$, the model predicts the 75th percentile of $Y$, conditional on covariates. Covariate effects are thus interpretable as how predictors shift the entire distribution of $Y$rather than just its mean. This is particularly useful when the response distribution is skewed, bounded and heavy-tailed. Unlike a classical linear model, which forces a constant marginal effect across the entire distribution, the CDF-beta model allows effects to vary across quantiles and does so automatically, without the need to run separate quantile regressions. The marginal effect of $X_j$on $y_i$is obtained by

$$\frac{\partial y_i}{\partial X_j} = q_Y(\mu_i)\,\frac{\partial \mu_i}{\partial X_j} = \beta_j\,\mu_i(1-\mu_i)\,q_Y(\mu_i) = g_i\beta_j,$$

where $q_Y(\mu_i)$ denotes the derivative of the quantile function, and $g_i = \mu_i(1-\mu_i)\,q_Y(\mu_i)$ is a multiplicative weight that varies across observations. This formulation indicates that, conditional on all other variables, a one-unit increase in $X_j$changes $y_i$by $g_i\beta_j$. Because $g_i$depends on $\mu_i$, the effect of $X_j$is inherently non-constant, illustrating the heteroscedastic nature of beta regression. However, the $q_Y(\mu_i)$ can be estimated using numerical derivative of $\left(\hat{\mu}_i, \hat{Q}_Y(\hat{\mu}_i\,)\right)$ that are available in R-software, such as "gradient" function in "pracma" package (R-software, R Core Team, 2026).

## 4.2 Precision model

In CDF-Based beta regression, the precision parameter $\phi$ need not be assumed constant; instead, it can be modelled through a dispersion model of the form $h(\phi_i) = Z_i^{\top}\gamma$, where the most common choice of link function is the logarithmic link,

$$\log(\phi_i) = Z_i^{\top}\gamma.$$

The vector $Z_i$may coincide with the covariates in the mean model $X_i$, or it may include a different set of predictors, and the coefficients $\gamma$ govern the behaviour of the dispersion (precision) component. This model captures systematic variability around the conditional mean and therefore plays a central role in characterizing heteroscedasticity in beta regression. The precision $\log\,(\phi_i) = Z_i^{\top}\gamma$ governs the dispersion of the transformed response $\mu_i$. Because the variance of a beta distribution satisfies $\mu_i(1-\mu_i)/(1+\phi_i)$ , larger values of $\phi_i$correspond to tighter clustering of observations around the predicted percentile, and smaller values correspond to greater spread. Thus, a positive $\gamma_k$implies that increasing covariate $Z_k$reduces variability of the response distribution. A negative $\gamma_k$implies that $Z_k$is associated with greater dispersion, reflecting heteroscedasticity in the data. This provides a transparent interpretation of variance heterogeneity, a task much harder to achieve with classical WLS/GLS models, where users must specify the variance function explicitly.

## 4.3 Bootstrap Inference

Given the nature of two-stage estimator, the nonparametric bootstrap offers a natural and robust method for inferential framework. Each bootstrap replication (R) proceeds by

1) resampling entire observations of $(Y_i, X_i, Z_i)$ with replacement,
2) recomputing the empirical CDF and the transformed values $U_i^{*}$,
3) refitting the beta mean and precision models,
4) generating U-scale prediction intervals via beta distribution quantiles,
5) mapping interval bounds back to the original scale using the empirical quantile function.

This procedure automatically captures the uncertainty arising from both the CDF estimation and the beta likelihood where it yields bootstrap standard errors, percentile-based confidence intervals for $(\beta, \gamma)$, and prediction intervals on the original response scale with correct coverage properties. Because bootstrap inference directly mirrors the complete estimator, it avoids the technical complications of analytic variance derivation while remaining fully consistent with the asymptotic behaviour of estimator (Efron and Tibshirani, 1994) and (Abdel-Karim, 2023). Nonparametric bootstrap algorithm to obtain standard errors, confidence interval and prediction intervals for the CDF-based beta can be described follows:

1. Fit on the original sample,
   - Compute CDF values $U_i = c((1, ((2:(n-2)) - 0.5), (n-1))/n$ (average ranks for ties), note
   - Fit Beta regression with mean link $g(\mu_i) = X\beta$ (logit) and precision link $h(\phi_i) = \log(\phi_i) = Z\gamma$.
   - Store $\hat{\theta} = (\hat{\beta}^{\mathrm{T}}, \hat{\gamma}^{\mathrm{T}})^{\mathrm{T}}$.
2. For $b = 1, \ldots, B$ bootstrap replications,
   - Resample rows by drawing $n$ indices with replacement $1, \ldots, n$ to form $(Y_i^{*(b)}, X_i^{*(b)}, Z_i^{*(b)})$.
   - Recompute CDF on the bootstrap sample $U_i^{*(b)} = c\big((1, \big(2:(n-2)\big) - 0.5\big), (n-1)\big)/n$,
   - Refit Beta regression on $U^{*(b)}$using the same links (logit for $\mu$, log for $\phi$); extract $\hat{\theta}^{*(b)}$.
3. Prediction on the original scale for any design matrix $X^{**}, Z^{**}$
   - Compute $\hat{\mu}^{*(b)} = \mathrm{logit}^{-1}(X^{**\mathrm{T}}\hat{\beta}^{*(b)})$ and $\hat{\phi}^{*(b)} = \exp\,(Z^{**\mathrm{T}}\hat{\gamma}^{*(b)})$.
   - From beta shape parameters $a^{*(b)} = \hat{\mu}^{*(b)}$ and $\hat{\phi}^{*(b)}$, $b^{*(b)} = (1 - \hat{\mu}^{*(b)})\hat{\phi}^{*(b)}$.
   - For a $1 - \alpha$ prediction interval (PI):
     - $u_{L^{*(b)}} = \mathrm{qbeta}\left(\frac{\alpha}{2}, a^{*(b)}, b^{*(b)}\right)$ and $u_{U^{*(b)}} = \mathrm{qbeta}\left(1 - \frac{\alpha}{2}, a^{*(b)}, b^{*(b)}\right)$, L: lower, U: upper.
     - Map to original scale by your empirical quantile function $Q_Y(\cdot)$ where $y_L^{*(b)} = Q_Y\left(u_L^{*(b)}\right)$, and $y_U^{*(b)} = Q_Y(u_U^{*(b)})$.
4. Summaries empirical standard error: error (SD) of $\hat{\theta}^{*(b)}_b = 1^B$, confidence intervals for coefficients: percentile or BCa from $\hat{\theta}^{*(b)}$ and prediction intervals on original scale by obtaining percentile bands of $y_L^{*(b)}, y_U^{*(b)}$or use the percentiles of $Q_Y(\mathrm{qbeta}(\cdot))$.

Note that, these bootstrap repeats both stages, so it captures uncertainty from the CDF transformation and the Beta MLE; it avoids explicitly deriving the variance contributed by the smoothed CDF and mapping predictions back via $Q_Y(\cdot)$ preserves interpretability on the original scale.

## 5 Simulation

To evaluate the performance of the proposed CDF-based Beta regression method. A comprehensive Monte-Carlo simulation study was conducted. The objective was to compare the distributional accuracy, quantile performance, prediction-interval calibration, and point-prediction error of the proposed method against commonly used alternatives, including log-transformed linear regression, reciprocal-transformed regression, ordinary least squares (OLS), weighted least squares (WLS), and generalized least squares (GLS).

The simulation design intentionally includes settings where classical approaches typically perform well (light-tailed, mildly heteroscedastic errors) and settings where they fail (strong heteroscedasticity, heavy-tailed errors, nonlinear mean structure). This allows us to demonstrate clearly under which conditions the proposed CDF-beta approach provides superior modelling of the full conditional distribution.

For each scenario, a dataset of size $n$ was generated using the generic model

$$Y_i = \mu_i + \sigma_i \varepsilon_i,$$

where both $\mu_i$and $\sigma_i$depend on covariates, and $\varepsilon_i$follows light-tailed or heavy-tailed distributions depending on the scenario. Three simulation conditions were imposed, each designed to reflect increasing levels of complexity and deviation from classical linear model assumptions. A total of $R = 1000$ Monte-Carlo replications was used for each condition. First scenario is S1 (mild heteroscedasticity, light-tailed errors) that represents a setting favourable to classical linear regression were covariates and mean function

$$x_1 \sim Un(-1,2), x_2 \sim N(0,1) \text{ and } \mu_i = 1 + 0.8x_{1i} - 0.5x_{2i},$$

variance function and error

$$\sigma_i = \exp(0.2 + 0.2x_{1i}), \text{and } \varepsilon_i \sim N(0,1)$$

Second scenario is S2 (strong heteroscedasticity, heavy-tailed errors ($t$-distribution with $df = 3$)) that introduces severe violations of the constant variance and normality assumptions were covariates, mean function

$$x_1 \sim Un(-1,2), x_2 \sim N(0,1) \text{ and } \mu_i = 1 + 0.8x_{1i} - 0.5x_{2i},$$

variance function and error

$$\sigma_i = \exp(0.6 + 0.8x_{1i}), \text{and } \varepsilon_i \sim t(df = 3)$$

To evaluate robustness when data exhibit both high heteroscedasticity and extreme non-normality conditions where log, reciprocal transforms, WLS, and GLS typically become unstable. Third scenario is S3 (nonlinear mean, strong heteroscedasticity, heavy-tailed errors) that combining nonlinear regression structure with extreme distributional irregularities assumptions were covariates, mean function

$$x_1 \sim Un(-1,2), x_2 \sim N(0,1) \text{ and } \mu_i = 1 + 2\sin(x_{1i}) + 0.5x_{1i}x_{2i},$$

variance function and error

$$\sigma_i = \exp(0.8 + 0.9x_{1i}), \text{and } \varepsilon_i \sim t(df = 3)$$

To test the ability of the CDF-beta method to model the entire distribution when the mean and dispersion structures are both highly nonlinear and the error distribution is far from normal.

To assess the performance of the proposed CDF-based beta regression model and the competing methods, we employ a set of metrics that evaluate both distributional accuracy and point prediction accuracy. The following metrics are used throughout the simulation study. Integrated CDF Error (ICDFE) quantifies the overall discrepancy between the true cumulative distribution function $F_Y(y)$ and the predicted CDF $\hat{F}_Y(y)$ across the entire support of the response variable

$$\text{ICDFE} = \int_0^1 \left[F_Y(y) - \hat{F}_Y(y)\right]^2 \, dy$$

ICDFE metric directly measures the accuracy of the estimated conditional distribution rather than just moments; see, Mansouri et al. (2024).

Kolmogorov–Smirnov Distance (KS) represents the maximum vertical difference between the true and predicted CDFs

$$\text{KS} = \sup_{y} \mid F_Y(y) - \hat{F}_Y(y) \mid$$

A KS value of zero indicates perfect alignment between the true and predicted distributions, while larger values indicate greater distributional mismatch; see, Mansouri et al. (2024).

Prediction interval coverage assesses the calibration of predictive uncertainty. For lower and upper predicted bounds $Y_L, Y_U$, coverage is computed as

$$\text{Coverage} = \frac{1}{n}\sum_{i=1}^{n} I\left( y_{L,i} \leq y_i \leq y_{U,i} \right)$$

ideal performance corresponds to coverage near the target levels (80% or 95%) and $I$ is the indicator function. Under coverage indicates intervals that are too narrow, while over coverage indicates overly conservative intervals.

RMSE measures point prediction accuracy on the original scale of the response variable

$$\text{RMSE} = \sqrt{\frac{1}{n}\sum_{i=1}^{n}(y_i - \hat{y}_i)^2}$$

Lower RMSE values indicate better mean-based prediction performance. Each simulated dataset was fitted using CDF-beta regression by transforms $Y$ to $U = F(Y)$, fits $\beta(\mu, \phi)$ regression with mean and precision models and maps predictions back using the empirical quantile function $\hat{Q}(\hat{\mu}_i)$.

For other methods, log transformed linear regression, reciprocal transformed regression, OLS, WLS using estimated variance weights, GLS with a variance-power structure. To ensure valid computations, log and reciprocal models were fitted with small positive offsets when needed (e.g., to avoid log (0)or division by zero). The performance of the proposed CDF-based beta regression was evaluated across the three simulation conditions described above. For each condition, results were averaged over $R = 1000$ Monte-Carlo replications. Performance was assessed using distribution level metrics, integrated CDF error (ICDFE), Kolmogorov–Smirnov (KS) statistic, and prediction interval coverage as well as the traditional root mean squared error (RMSE). Lower values of ICDFE and KS indicate better distributional accuracy, whereas coverage is evaluated by closeness to the target nominal level (80% or 95%).

Table 1 shows result for scenario S1, where model assumptions are only mildly violated. Under this setting, classical methods (OLS, WLS, GLS) perform reasonably well, yet the CDF-based beta regression still achieves the best distributional accuracy for all sample sizes. For example, at n = 25, the Beta model attains an ICDFE of 0.0025, far below GLS (0.0129), OLS (0.0122), and WLS (0.0120). The KS statistic for the proposed model is considerably smaller (0.0317) than those of GLS (0.1207) and OLS (0.1177), indicating that it follows the true conditional distribution much more closely. The prediction intervals from the beta model also remain well calibrated. Even with a small sample size (n = 25), the 80% coverage is 0.7994 and the 95% coverage is 0.9640, both very close to their nominal levels. These results are comparable to GLS and OLS, suggesting that the model provides reliable uncertainty estimates even in relatively simple settings. Regarding predictive accuracy, RMSE values are quite similar across all methods. Although OLS achieves the smallest RMSE in case of (1.2872 at n = 25), the CDF-beta model remains highly competitive (1.3333 at n = 25), offering strong predictive performance alongside its improved distributional fit.

Table 1. Results of 1000 Monte-Carlo simulations for scenario S1 (mild heteroscedasticity, light-tailed errors) ) in terms of ICDFE, KS, RMSE, coverage at 80% and 90%.

| | | Methods | | | | | |
|---|---|---|---|---|---|---|---|
| n | Metric | Recip | Log | OLS | GLS | WLS | Beta |
| 25 | ICDFE | 0.8731 | 0.3617 | 0.0122 | 0.0129 | 0.0120 | 0.0025 |
| | KS | 0.8205 | 0.3731 | 0.1177 | 0.1207 | 0.1170 | 0.0317 |
| | RMSE | 21.6916 | 3.5433 | 1.2872 | 1.2958 | 1.2961 | 1.3333 |
| | Cov80 | - | - | 0.7862 | 0.7869 | 0.7890 | 0.7994 |
| | Cov95 | - | - | 0.9629 | 0.9617 | 0.9640 | 0.9640 |
| 50 | ICDFE | 1.0351 | 0.3781 | 0.0062 | 0.0064 | 0.0062 | 0.0015 |
| | KS | 0.8186 | 0.3335 | 0.0826 | 0.0838 | 0.0822 | 0.0248 |
| | RMSE | 34.0067 | 2.2372 | 1.3423 | 1.3449 | 1.3460 | 1.3586 |
| | Cov80 | - | - | 0.7925 | 0.7932 | 0.7929 | 0.7999 |
| | Cov95 | - | - | 0.9554 | 0.9546 | 0.9555 | 0.9558 |
| 75 | ICDFE | 50.5294 | 1.8915 | 1.3582 | 1.3598 | 1.3606 | 0.0013 |
| | KS | 0.8212 | 0.3083 | 0.0672 | 0.0682 | 0.0670 | 0.0238 |
| | RMSE | 50.5294 | 1.8915 | 1.3582 | 1.3598 | 1.3606 | 1.3799 |
| | Cov80 | - | - | 0.7958 | 0.7963 | 0.7965 | 0.8019 |
| | Cov95 | - | - | 0.9525 | 0.9525 | 0.9528 | 0.9516 |
| 100 | ICDFE | 1.2047 | 0.3455 | 0.0031 | 0.0032 | 0.0031 | 0.0011 |
| | KS | 0.8225 | 0.2900 | 0.0578 | 0.0586 | 0.0577 | 0.0223 |
| | RMSE | 48.7062 | 1.7458 | 1.3686 | 1.3696 | 1.3703 | 1.3891 |
| | Cov80 | - | - | 0.7950 | 0.7949 | 0.7952 | 0.8013 |
| | Cov95 | - | - | 0.9522 | 0.9522 | 0.9524 | 0.9521 |

-unreliable coverage number

Table 2. Results of 1,000 Monte-Carlo simulations for scenario S2 (strong heteroscedasticity, heavy-tailed errors) in terms of ICDFE, KS, RMSE, coverage at 80% and 90%.

| | | Methods | | | | | |
|---|---|---|---|---|---|---|---|
| n | Metric | Reci | log | OLS | GLS | WLS | Beta |
| 25 | ICDFE | 5.0721 | 3.1178 | 0.1426 | 0.1590 | 0.1493 | 0.0206 |
| | KS | 0.7157 | 0.4371 | 0.1572 | 0.1695 | 0.1475 | 0.0569 |
| | RMSE | 50.5991 | 19.2038 | 5.8607 | 6.0592 | 6.1418 | 6.1513 |
| | Cov80 | - | - | 0.8121 | 0.8158 | 0.8146 | 0.8010 |
| | Cov95 | - | - | 0.9477 | 0.9450 | 0.9441 | 0.9538 |
| 50 | ICDFE | 7.9317 | 4.0744 | 0.1481 | 0.1389 | 0.1301 | 0.0174 |
| | KS | 0.7275 | 0.4197 | 0.1296 | 0.1428 | 0.1198 | 0.0518 |
| | RMSE | 366.6827 | 12.6059 | 6.3889 | 6.4847 | 6.4798 | 6.5052 |
| | Cov80 | - | - | 0.8356 | 0.8381 | 0.8411 | 0.8019 |
| | Cov95 | - | - | 0.9428 | 0.9418 | 0.9416 | 0.9600 |
| 75 | ICDFE | 9.4173 | 4.5357 | 0.1167 | 0.1180 | 0.1195 | 0.0162 |
| | KS | 0.7298 | 0.4120 | 0.1162 | 0.1293 | 0.1079 | 0.0496 |
| | RMSE | 590.3935 | 10.9191 | 6.3878 | 6.4470 | 6.4837 | 6.5019 |
| | Cov80 | - | - | 0.7958 | 0.7963 | 0.7965 | 0.8029 |
| | Cov95 | - | - | 0.9428 | 0.9425 | 0.9425 | 0.9553 |
| 100 | ICDFE | 11.5017 | 5.2635 | 0.1391 | 0.1441 | 0.1555 | 0.0157 |
| | KS | 0.7346 | 0.4052 | 0.1113 | 0.1235 | 0.1062 | 0.0476 |
| | RMSE | 80.6026 | 11.1760 | 6.7595 | 6.8114 | 6.8375 | 6.8555 |
| | Cov80 | - | - | 0.8531 | 0.8527 | 0.8559 | 0.8028 |
| | Cov95 | - | - | 0.9450 | 0.9455 | 0.9445 | 0.9596 |

-unreliable coverage number

Table 2 shows the simulation results under strong heteroscedasticity and heavy-tailed errors. The CDF-based beta regression is markedly doing well in most metrics. For example, at n = 25, Beta regression achieves an ICDFE of 0.0206, compared with 0.1590 (GLS), 3.1178 (log), 5.0721 (reciprocal), and 0.1426 (OLS). Similarly, the KS statistic at n = 25 is 0.0569 for the beta model, far below GLS (0.1695), log (0.4371), and reciprocal (0.7157), indicating much closer alignment with the true distribution. Prediction interval coverage further illustrates the advantage. At n = 100, the Beta model achieves 95% coverage of 0.9596, closely matching nominal levels, whereas GLS, OLS, and WLS show lower or unstable coverage.
Transformation-based models cannot report coverage, due to invalid back-transformation of prediction intervals under non-normality. RMSE values increase substantially for all methods due to heavy-tailed errors, with the CDF-beta model remaining competitive (e.g., 6.1513 at n = 25) while transformation-based models such as the log and reciprocal regressions perform extremely poorly, reaching RMSE values as high as 19.2038 and 50.5991, respectively.

Table 3. Results of 1000 Monte-Carlo simulations for scenario S3 (nonlinear Mean, strong heteroscedasticity, heavy-tailed errors) in terms of ICDFE, KS, RMSE, coverage at 80% and 90%.

| | | Methods | | | | | |
|---|---|---|---|---|---|---|---|
| n | Metric | Reci | log | OLS | GLS | WLS | Beta |
| 25 | ICDFE | 7.3052 | 4.4659 | 0.2104 | 0.2904 | 0.2090 | 0.0363 |
| | KS | 0.6871 | 0.4409 | 0.1626 | 0.1783 | 0.1484 | 0.0621 |
| | RMSE | 198.6876 | 27.5385 | 8.1889 | 8.5297 | 8.4356 | 8.5680 |
| | Cov80 | - | - | 0.8106 | 0.8157 | 0.8161 | 0.7991 |
| | Cov95 | - | - | 0.9472 | 0.9441 | 0.9445 | 0.9544 |
| 50 | ICDFE | 11.1498 | 6.0503 | 0.2081 | 0.2042 | 0.2050 | 0.0330 |
| | KS | 0.7015 | 0.4277 | 0.1320 | 0.1462 | 0.1203 | 0.0584 |
| | RMSE | 56.1624 | 18.6682 | 8.8301 | 8.9812 | 9.0356 | 9.0473 |
| | Cov80 | - | - | 0.8325 | 0.8354 | 0.8383 | 0.7989 |
| | Cov95 | - | - | 0.9428 | 0.9425 | 0.9425 | 0.9581 |
| 75 | ICDFE | 13.7551 | 6.7125 | 0.2283 | 0.2228 | 0.1681 | 0.0317 |
| | KS | 0.7094 | 0.4167 | 0.1193 | 0.1326 | 0.1074 | 0.0571 |
| | RMSE | 140.2100 | 16.2585 | 9.1082 | 9.2216 | 9.2090 | 9.2495 |
| | Cov80 | - | - | 0.8440 | 0.8447 | 0.8477 | 0.7986 |
| | Cov95 | - | - | 0.9450 | 0.9431 | 0.9445 | 0.9549 |
| 100 | ICDFE | 15.5504 | 7.0894 | 0.1510 | 0.1666 | 0.1503 | 0.0323 |
| | KS | 0.7131 | 0.4097 | 0.1088 | 0.1235 | 0.0994 | 0.0569 |
| | RMSE | 111.6167 | 15.0528 | 9.1188 | 9.2009 | 9.2022 | 9.2545 |
| | Cov80 | - | - | 0.8481 | 0.8490 | 0.8525 | 0.7978 |
| | Cov95 | - | - | 0.9445 | 0.9449 | 0.9435 | 0.9573 |

- unreliable coverage number

Table 3 shows the simulation that results for scenario (S3), combining nonlinear mean structure, strong heteroscedasticity, and heavy-tailed errors. Across all sample sizes, the CDF-based beta regression again provides the best distributional accuracy, achieving far smaller ICDFE and KS values than competing methods. For example, at n = 25, Beta regression attains an ICDFE of 0.0363, compared with 0.2104 (OLS) and 4.4659 (log), and a KS of 0.0621 versus 0.1626 (OLS) and 0.4409 (log). Regarding RMSE, all methods show inflated errors due to the nonlinear mean and heavy-tailed noise, yet the Beta model remains competitive (e.g., 8.5680 at n = 25), while transformation-based models behave erratically, reaching 27.5385 (log) and 198.6876 (reciprocal) at the same sample size.
Overall, the simulation results confirm that CDF-beta is a robust and distribution centric regression framework in terms of ICDFE, KS, and coverage. As a result, it can provide accurate inference and reliable predictions under a wide range of data generating mechanisms, including the classical methods that break down.

# 6 Real data application

The empirical analysis uses the concrete compressive strength dataset that consists of 1030 observations and nine continuous variables, originally provided by the UCI Machine Learning Repository (https://archive.ics.uci.edu/dataset/165/concrete+compressive+strength). The dataset includes eight input features describing the material composition and curing age of concrete including cement, blast furnace slag, fly ash, water, superplasticizer, coarse and fine aggregate, and age in days, along with a continuous response variable measuring the compressive strength of concrete (MPa)**.** The aim of the regression analysis is to model and predict the compressive strength from the mixture components.

To evaluate the suitability of the OLS model for the concrete compressive strength data, we conducted a standard set of heteroscedasticity diagnostics. The initial OLS fit revealed clear signs of variance instability.

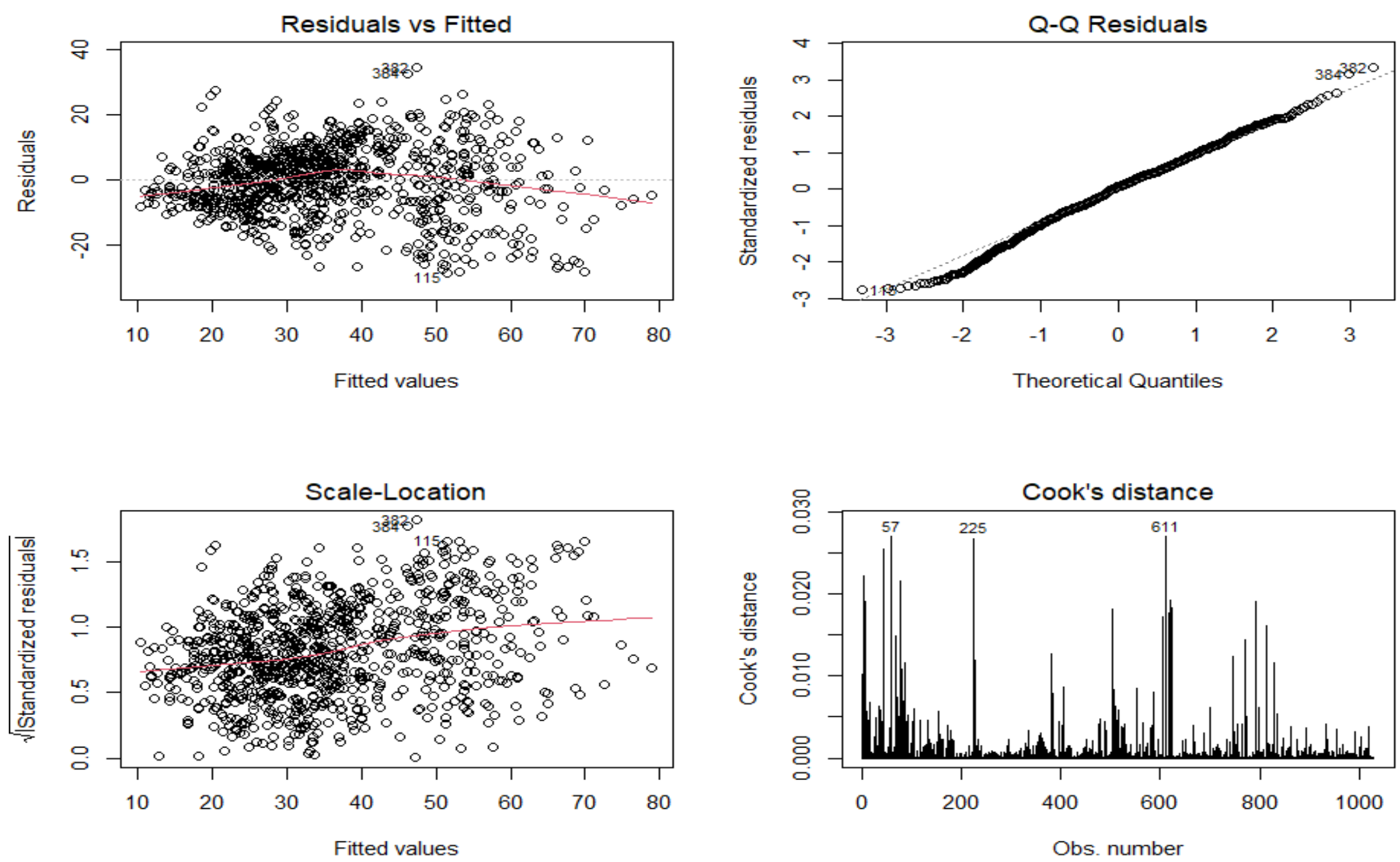

Figure 1. standard OLS diagnostic plots for concrete compressive strength data

In Figure 1, the residuals show a systematic funnel-shaped pattern, with the spread of residuals increasing at higher fitted values that indicating evidence of heteroscedasticity.

In QQ plot, the points deviate noticeably from the 45° reference line, especially in the tails. This pattern suggests that the residuals are not normally distributed and the data exhibit heavy tail distribution while the scale–location plot provides further evidence of heteroscedasticity. Where the LOESS curve slopes down, the variance of the residuals increases as the predicted compressive strength becomes larger. In addition, several data display large Cook's distances. A few cases (such as observations 57, 225, and 611) stand out markedly from the rest, indicating that they exert a strong influence on the OLS estimates. This makes the model unstable and highly sensitive to outliers.

Formal statistical testing supports these visual findings. Because the Breusch–Pagan test is highly significant (BP = 116.85, $p < 0.001$), the presence of heteroscedasticity in the model is confirmed. This gives evidence that the residual variance depends on one or more predictors rather than remaining constant. However, no series multicollinearity found in the data where the variance inflation factor (VIF) for all variable between (1.5, 7). These findings suggest that the dataset suffers from both nonnormality and heteroscedasticity and motivate the need of suitable model.

To address these issues, we applied the two-stage CDF-based beta regression framework to the dataset. We then assessed how well the model fit by using DHARMa's simulated residuals (Hartig (2024) and R Core Team (2026)), which are expected to follow a uniform distribution when the model is correctly specified.

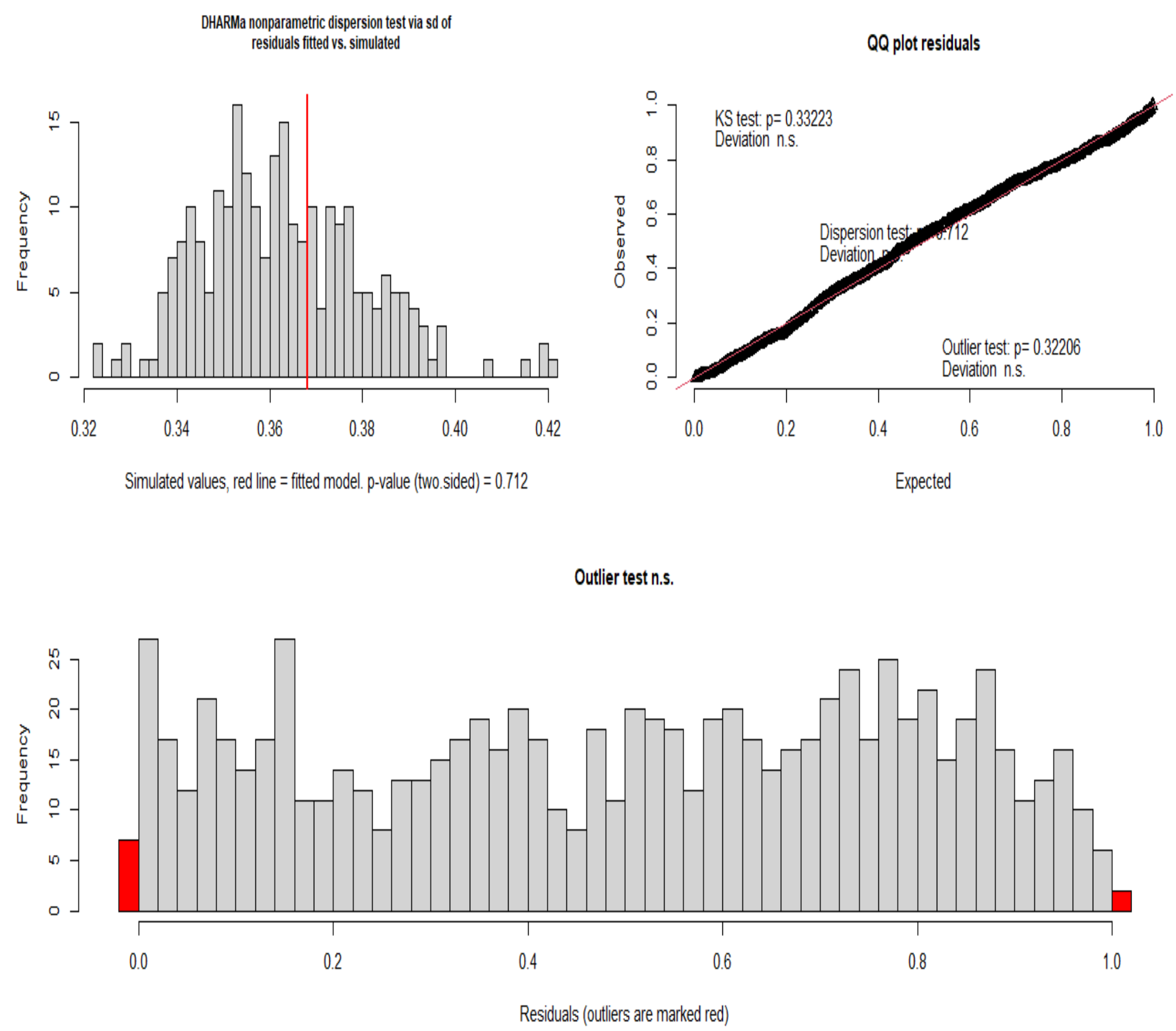

Figure 2. DHARMa diagnostic plots for the CDF-Based beta regression Model using concrete compressive strength data

In Figure 2, the QQ plot of DHARMa residuals showed close alignment with the theoretical uniform (0,1) distribution. The Kolmogorov–Smirnov test for residual uniformity was nonsignificant ($p = 0.332 > 0.05$). The nonparametric dispersion test showed no evidence of over or under dispersion ($p = 0.712 > 0.05$) and the outlier test indicated no excess of extreme residuals beyond model expectations ($p = 0.322 > 0.05$). These diagnostics collectively support that the CDF-based beta model is well modelled the data. The interpretable predications in the nature units of the response can be obtained from fitted percentiles $\hat{\mu}_i$ where it is mapped back to the original outcome scale using the empirical quantile function $\hat{y}_i = Q_Y(\hat{\mu}_i)$.

Table 4 reports the estimated results of the model. From the table, the pseudo $R^2$ value of 0.571 suggests that the model explains a good proportion of the variation when the outcome is expressed on the percentile scale. In the mean component of the model, all variables have positive coefficients except water. This means that as these variables increase, the predicted percentile of compressive strength shifts upward. In contrast, water has a negative and statistically significant effect ($-0.0159$, p < 0.001). As a result, the higher water content lowers the predicted percentile of strength. This aligns with the well-known relationship that increasing water content generally reduces concrete strength. Since the coefficients are estimated on the logit percentile scale, their marginal effects on the expected percentile are nonlinear ($(\beta_j\, \mu_i(1-\mu_i))$). Consequently, the impact of each predictor is strongest around the middle of the distribution and becomes smaller toward the lower and upper tails. Note that, coarse and fine did not show statistical effect, so they are not reported in Table 4.
For illustration, at the mid-distribution ($\mu_i \approx 0.5$), a one-unit increase in cement raises the expected percentile by about ($0.25 \times 0.0084 \approx 0.0021$ (0.21 percentile points) while water lowers it by $0.25 \times 0.0159 \approx 0.0040$(0.40 percentile points). Mapping fitted percentiles back to the original units is achieved via the quantile function $\hat{y}_i = Q_Y(\hat{\mu}_i)$ that enable interpretation directly in MPa when desired.
Because the precision coefficient for slag is negative ($-0.0017$, p < 0.001), the higher slag content leads to greater variability around the predicted percentile. In contrast, water has a small positive effect on precision (0.0033, p = 0.091>0.05), though this result is only marginal and not statistically significant at the 5% level.

Table4. Estimated coefficients of the concrete strength data using CDF-Based beta regression

| component | term | estimate | std.error | statistic | p.value |
|---|---|---|---|---|---|
| mean | (Intercept) | -0.9136 | 0.3251 | -2.8101 | 0.0050 |
| mean | cement | 0.0084 | 0.0003 | 24.4372 | 0.0000 |
| mean | slag | 0.0067 | 0.0004 | 16.5209 | 0.0000 |
| mean | fly_ash | 0.0057 | 0.0006 | 9.1537 | 0.0000 |
| mean | water | -0.0159 | 0.0016 | -9.7308 | 0.0000 |
| mean | superplasticizer | 0.0206 | 0.0070 | 2.9482 | 0.0032 |
| mean | age_days | 0.0124 | 0.0005 | 25.3583 | 0.0000 |
| precision | (Intercept) | 1.2883 | 0.3502 | 3.6788 | 0.0002 |
| precision | slag | -0.0017 | 0.0005 | -3.7469 | 0.0002 |
| precision | water | 0.0033 | 0.0019 | 1.6919 | 0.0907 |
| Pseudo R-squared | | 0.571 | | | |

Table5. Performance comparison of OLS, LOG, and WL with CDF-Based Beta

| Method | RMSE | Cov80 | Cov95 |
|---|---|---|---|
| OLS (original scale) | 10.520 | 0.771 | 0.941 |
| Log-OLS (Duan smearing) | 13.268 | 0.830 | 0.956 |
| WLS (variance-power weights) | 10.796 | 0.1067 | 0.169 |
| Two-stage CDF-Based Beta | 10.544 | 0.815 | 0.953 |

Table 5 reports the performance of the proposed approach (CDF-based beta) with other competitors (OLS, LOG and WL) that applied to the concrete compressive strength data in terms of RMSA, Cov80 and Cov95. In general, the results show that the two stage CDF-based beta regression provides the most balanced and reliable performance across all evaluation metrics. In terms of prediction accuracy, RMSE values for all methods fall within a narrow band (10.52, 10.80), with the CDF-based Beta method (RMSE = 10.54) closely matching OLS (10.52) and outperforming WLS and classical beta regression. However, the clearest differences emerge in interval calibration. The 95% prediction interval coverage for the CDF-based beta model (0.953) is the nearest to nominal among all methods, OLS (0.942) and Log-OLS with Duan smearing (0.956). on the other hand, the variance-power WLS model performs extremely poorly, with severe under-coverage at both 80% (0.107) and 95% (0.170). This is confirming instability of variance-function–based corrections under heteroscedasticity and heavy-tailed errors. These results highlight that the CDF-based beta method is the most robust and reliable method for this dataset where it achieves near OLS level point accuracy while delivering substantially better uncertainty calibration.

# 7 Conclusion

This study proposed a two-stage CDF-based beta regression framework as a distribution-centric alternative for modelling continuous outcomes that exhibit heteroscedasticity, skewness, or non-normality. Instead of relying on traditional assumptions, the approach first transforms the response variable using a smoothed empirical CDF and then models the resulting percentiles with a beta distribution. By working directly with the distribution of the data, the method naturally accounts for mean-variance dependent and other irregularities, without the need for arbitrary transformations or predefined variance structures.

The simulation results showed clear advantages across scenarios with mild, moderate, and severe violations of standard model assumptions. The CDF-based beta approach consistently delivered better distributional accuracy, achieving the lowest ICDFE and KS values while it maintained RMSE performance comparable to OLS and GLS. Consequently, it demonstrated that improved distributional fit does not come at the expense of predictive accuracy.

When we applied the proposed method to the concrete compressive strength data, the benefits became clear in practice. The traditional OLS model showed several problems in its diagnostics, including clear heteroscedasticity, non-normal residuals, and some influential observations that could affect the results. In contrast, the CDF-based beta regression performed much more smoothly. Its residuals behaved as expected (uniform under DHARMa diagnostics) where there was no indication of dispersion issues. Although all models achieved similar RMSE values, the real difference appeared in prediction accuracy where the CDF-based beta regression produced prediction intervals with coverage probabilities very close to the nominal 80% and 95% levels, while OLS, Log-transformation, and WLS did not perform as well in this respect. Overall, the results demonstrate the practical advantage of the proposed approach for this dataset.

In short, the two-stage CDF-based beta framework provides a clear, practical, and intuitive way to model complex data. It remains stable in real applications and produces reliable statistical inference and well-calibrated predictions, even when the assumptions of traditional linear models are violated. This approach may be expanded to handle multilevel or longitudinal data or further refined by exploring alternative smoothing techniques.

**Appendix**

Let $Y_1, \dots, Y_n$ be i.i.d. observations from a distribution with cumulative distribution function $F(y)$. The empirical distribution function (EDF) is defined as

$$\hat{F}_n(y) = \frac{1}{n}\sum_{i=1}^{n} I(Y_i \le y).$$

It is well known that $\hat{F}_n(y)$ is a consistent estimator of $F$ and

$$\sqrt{n}\left(\hat{F}_n(y) - F(y)\right) \xrightarrow{d} \mathcal{N}(0, F(y)(1 - F(y))),$$

(Serfling, 1980).

**Theorem A.1**

Let $F_{C,i}(y)$ be the smoothed empirical CDF of $Y$. Then $F_{C,i}(y)$is a $\sqrt{n}$-consistent estimator of the true CDF $\left(F(y)\right)$and satisfies

$$\sqrt{n}\,(F_{C,i}(y) - F(y)) \xrightarrow{d} N(0, \sigma^2_{\text{smooth}}(y)),$$

where

$$0 < \sigma^2_{\text{smooth}}(y) < F(y)\{1 - F(y)\}.$$

Thus, $F_{C,i}(y)$is $\sqrt{n}$-consistent.

*Proof*
Because $F_{C,i}(y)$is a smoothed version of the empirical distribution function, it differs from the EDF by $o_p(n^{-1/2})$. Since the classical EDF satisfies

$$\sqrt{n}(F_n(y) - F(y)) \Rightarrow N(0, F(y)(1 - F(y))),$$

the smoothed estimator inherits the same convergence rate but with reduced variance.

**Theorem A.2**
Consider the transformed response $U_i = F_{C,i}(Y_i)$and the beta regression model

$$U_i \sim \text{Beta}(\mu_i, \phi_i),$$

With mean and precision models

$$\text{logit}(\mu_i) = X_i^{\text{T}}\beta, \text{and } \log(\phi_i) = Z_i^{\text{T}}\gamma.$$

If $F_{C,i}(y)$is $\sqrt{n}$-consistent, then the maximum likelihood estimator

$$\hat{\theta} = \left(\hat{\beta}^T, \hat{\gamma}^T\right)^T$$

satisfies

$$\sqrt{n}\,(\hat{\theta} - \theta_0) \xrightarrow{d} N(0, I^{-1}(\theta_0)),$$

meaning the CDF-based beta estimator has the same first-order asymptotic distribution as the classical beta-regression MLE.
*Proof*
By theorem A.1,

$$U_i = F(Y_i) + O_p(n^{-1/2}),$$

the transformation error is of order $n^{-1/2}$. A Taylor expansion of the beta-likelihood score gives

$$S(\theta; U_i) = S(\theta; F(Y_i)) + O_p(n^{-1/2}),$$

the perturbation contributes only a second-order term when summed over all $i$. Hence the dominant $O(n)$likelihood term is unaffected, and the estimator retains the classical $\sqrt{n}$-normal asymptotic

$$\sqrt{n}(\hat{\theta} - \theta_0) \Rightarrow N(0, I^{-1}(\theta_0)).$$